\documentclass{emulateapj}
\usepackage{epstopdf}

\newcommand{\ldl}{$\lambda/{\Delta}{\lambda}$}
\newcommand{\teff}{T$_{eff}$}

\newcommand{\lii}{\ion{Li}{1}}
\newcommand{\ki}{\ion{K}{1}}
\newcommand{\nai}{\ion{Na}{1}}

\newcommand{\meth}{CH$_4$}
\newcommand{\water}{H$_2$O}
\newcommand{\wat}{H$_2$O}
\newcommand{\name}{SDSS~J080531.84+481233.0}
\newcommand{\namesh}{SDSS~J0805+4812}

\slugcomment{Accepted for publication to AJ}

\shorttitle{The Unresolved Binary \namesh}
\shortauthors{Burgasser}

\begin{document}

\title{SDSS~J080531.84+481233.0: An Unresolved L Dwarf/T Dwarf Binary}

\author{Adam J.\ Burgasser\altaffilmark{1}}

\affil{Massachusetts Institute of Technology, Kavli Institute for Astrophysics and Space Research,
Building 37, Room 664B, 77 Massachusetts Avenue, Cambridge, MA 02139, USA; ajb@mit.edu}

\altaffiltext{1}{Visiting Astronomer at the Infrared Telescope Facility, which is operated by
the University of Hawaii under Cooperative Agreement NCC 5-538 with the National Aeronautics
and Space Administration, Office of Space Science, Planetary Astronomy Program.}

\begin{abstract}
SDSS~J080531.84+481233.0 is a 
peculiar L-type dwarf that exhibits unusually
blue near-infrared and mid-infrared colors and 
divergent optical (L4) and near-infrared
(L9.5) spectral classifications.  These peculiar spectral traits have been variously attributed to condensate cloud effects or subsolar metallicity.
Here I present an improved near-infrared spectrum of this source which  
further demonstrates the presence of
weak CH$_4$ absorption at 1.6 $\micron$ but no corresponding band at 2.2~$\micron$.  
It is shown that these features can be collectively reproduced
by the combined light spectrum of a binary with 
L4.5 and T5 components, as deduced by 
spectral template matching.
Thus, SDSS~J080531.84+481233.0 appears to be a new low-mass
binary straddling the L dwarf/T dwarf 
transition, an evolutionary phase for brown dwarfs
that remains poorly understood by current theoretical models. 
The case of
SDSS~J080531.84+481233.0 further illustrates how a select range of
L dwarf/T dwarf binaries could be identified
and characterized without the need for high angular resolution imaging
or radial velocity monitoring, 
potentially alleviating some of the detection biases and limitations
inherent to such techniques.
\end{abstract}

\keywords{
stars: binaries: general ---
stars: fundamental parameters ---
stars: individual ({\name}) ---
stars: low-mass, brown dwarfs
}

\section{Introduction}

Coeval systems, from binaries to dense clusters, are invaluable
resources for stellar studies.  By significantly reducing uncertainties
in distance, age and composition, 
multiple systems enable comparative analyses of atmospheric properties,
circumstellar environments, magnetic activity trends and angular momentum 
evolution.  Close binary systems also facilitate dynamical mass
measurements, as well as radius measurements for eclipsing systems.
The multiplicity characteristics of a coeval population provide 
critical constraints
for theories exploring stellar genesis, as well as 
the distribution of stellar and substellar
masses and the incidence of planetary systems throughout the Galaxy.

Multiple systems are of particular importance in studies of the
lowest-mass stars incapable of sustained core hydrogen fusion,
the so-called brown dwarfs.  The apparently low resolved binary fraction
of field brown dwarfs ($\sim$10--15\%; see \citealt{meppv} and references
therein) has been cited as evidence of mass-dependent multiple
formation (e.g., \citealt{bou06}), as predicted by some brown dwarf
formation models (e.g., \citealt{ste03}).  However, resolved imaging
studies provide only a lower limit to the true binary fraction,
and evidence from radial velocity studies (e.g., \citealt{max05})
and overluminous cluster members \citep{pin03,cha05,bou06} suggests 
a much higher total binary fraction, perhaps 25\% or more \citep{bas06,rei2252}.
This may prove to be a significant challenge for some brown dwarf formation
theories (e.g., \citealt{bat02}).

Unresolved multiples also play an important role in understanding
the transition between the two lowest-luminosity classes of known
brown dwarfs, the L dwarfs and T dwarfs \citep[and references therein]{kir05}.
This transition occurs when photospheric condensates, 
a dominant source of opacity in L dwarf atmospheres, disappear, resulting
in near-infrared spectral energy distributions that are blue and dominated
by molecular gas absorption, including {\meth}
\citep{tsu96,tsu99,bur99,cha00,all01}.
While condensate cloud models provide a physical
basis for this transition \citep{ack01,coo03,bur06},
they fail to explain its apparent rapidity,
as deduced by the small effective temperature ({\teff})
differential \citep{kir00,gol04,vrb04}
and apparent brightening at 1~$\micron$ \citep{dah02,tin03,vrb04} 
between late-type L dwarfs and mid-type T dwarfs.
Multiplicity effects may be partly responsible for these trends, 
particularly as the resolved binary fraction of L/T transition objects
is nearly twice that of other spectral types \citep{mehst2},
and can result in overestimated temperatures and surface fluxes
\citep{gol04,liu06}.  As the total binary fraction of L/T transition 
objects may be higher still (perhaps as high as 65\%; \citealt{meltbinary}),
interpretations of absolute brightness, color and {\teff}
trends across this important evolutionary phase for nearly all brown dwarfs
may be skewed.

Empirical constraints on the L/T transition 
can be made through the identification 
and characterization of binaries with components that
span this transition \citep{cru04,me0423,mehst2,liu06,rei2252}.  
One such system that may have been overlooked
is the peculiar L dwarf {\name} (hereafter {\namesh}; \citealt{haw02,kna04}),
identified in the Sloan Digital Sky Survey (hereafter SDSS; \citealt{yor00}).
This source has widely discrepant optical (L4; \citealt{haw02})
and near-infrared (L9.5$\pm$1.5; \citealt{kna04,chi06}) spectral types,
and unusually blue near-infrared colors ($J-K$ = 1.10$\pm$0.04;
\citealt{kna04})
compared to either L4 ($\langle J-K \rangle$ = 1.52) or 
L8-T0.5 dwarfs ($\langle J-K \rangle$ = 1.58--1.74; \citealt{vrb04}).
Its mid-infrared colors are also peculiar
\citep{gol04,kna04,leg06}.  These characteristics have 
been interpreted as resulting from a metal-poor 
photosphere or one with 
unusually thin photospheric condensate clouds 
\citep{kna04,gol04,leg06,fol07}.  
However, unresolved multiplicity may provide a better explanation for
the peculiar properties of this source.
In this article I present and analyze new low-resolution near-infrared
spectral data for {\namesh} that supports this hypothesis,
and demonstrate that this source is likely to be a binary 
with components straddling the L/T transition.
Spectral observations are described
in $\S$~2, including a detailed discussion of the unusual
features observed in these data.  
Analysis of these data in regard to its possible binary nature is
described in $\S$~3,
and the properties of the components inferred from this analysis is
discussed in $\S$~4.  Finally, the implications
of this study, including application of the technique used here
to identify and characterize brown dwarf binaries 
independent of angular resolution limitations, in briefly discussed 
in $\S$~5.

\section{Observations}

\subsection{Data Acquisition and Reduction}

Low resolution near-infrared
spectral data for {\namesh} were
obtained on 2006 December 24 (UT) using the SpeX spectrograph \citep{ray03}
mounted on the 3m NASA Infrared Telescope Facility (IRTF).
The conditions were clear with good seeing (0$\farcs$8 at $J$-band).
The 0$\farcs$5 slit was employed, providing 0.75--2.5~$\micron$
spectroscopy with resolution {\ldl} $\approx 120$
and dispersion across the chip of 20--30~{\AA}~pixel$^{-1}$.
To mitigate the effects of differential refraction, the slit was aligned
to the parallactic angle. Six exposures of 
120~s each were obtained 
in an ABBA dither pattern along the slit.
The A0~V star HD~71906 was observed immediately
afterward at a similar airmass ($z$ = 1.18) for flux calibration.
Internal flat field and argon arc lamps were also observed
for pixel response and wavelength calibration.

Data were reduced using the SpeXtool package version 3.4
\citep{cus04} using standard settings.
Raw science images were first
corrected for linearity, pair-wise subtracted, and divided by the
corresponding median-combined flat field image.  Spectra were optimally extracted using the
default settings for aperture and background source regions, and wavelength calibration
was determined from arc lamp and sky emission lines.  The multiple
spectral observations were then median-combined after scaling individual
spectra to match the highest signal-to-noise
observation.  Telluric and instrumental response corrections for the science data were determined
using the method outlined
in \citet{vac03}, with line shape kernels derived from the arc lines. 
Adjustments were made to the telluric spectra to compensate
for differing \ion{H}{1} line strengths in the observed A0~V spectrum
and pseudo-velocity shifts.
Final calibration was made by
multiplying the spectrum of {\namesh} by the telluric correction spectrum,
which includes instrumental response correction through the ratio of the observed A0~V spectrum
to a scaled, shifted and deconvolved Kurucz\footnote{\url{http://kurucz.harvard.edu/stars.html}.}
model spectrum of Vega. 

\subsection{The Spectrum of {\namesh}}

\begin{figure}
\epsscale{0.8}
\plotone{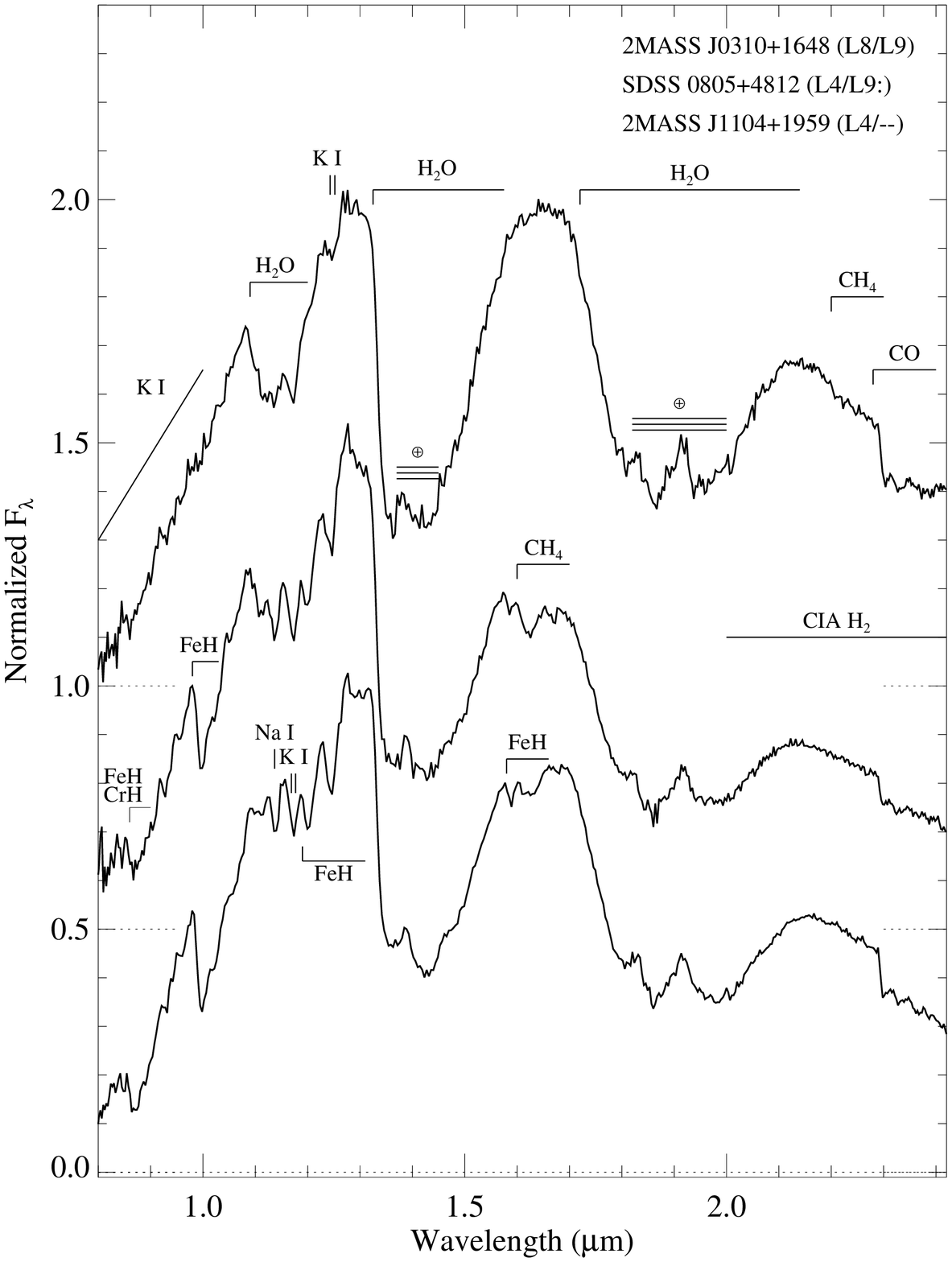}
\caption{Reduced SpeX prism spectrum for {\namesh} (center) compared
to equivalent data for 
the optically classified L4 2MASS~J1104+1959 (bottom) and the L8/L9
(optical/near-infrared type)
2MASS~J0310+1648 (top).  All three spectra
are normalized at their 1.25~$\micron$ flux peaks and offset by
constants (dotted lines).  Prominent features resolved by these
spectra are indicated. Note in particular the weak band of CH$_4$
at 1.6~$\micron$ in the spectrum of {\namesh}.
\label{fig_nirspec}}
\end{figure}

The reduced spectrum of {\namesh} is shown in Figure~\ref{fig_nirspec}, and
compared to equivalent SpeX prism data for the optically classified
L4 2MASS~J11040127+1959217 \citep[hereafter 2MASS~J1104+1959]{cru03}, and 
2MASS J03105986+1648155 \citep[hereafter 2MASS~J0310+1648]{kir00}
which is classified
L8 in the optical and L9 in the near-infrared \citep{geb02}.
The spectrum of {\namesh} is most similar to that of 2MASS~J1104+1959, 
based on their overall
spectral energy distributions, strong FeH absorption at 0.99~$\micron$, and prominent {\nai} and {\ki} lines in the 1.1--1.25 $\micron$ range.  
However, the 1.15 and 1.3~$\micron$ {\water} bands are 
clearly much stronger in the spectrum of {\namesh} 
but similar in strength to those
in the spectrum of 2MASS~J0310+1648.
Other spectral characteristics of {\namesh} are inconsistent with 
either of the comparison sources, such as the suppressed $K$-band
flux peak and weak CO absorption at 2.3~$\micron$.

The most unusual
feature observed in the spectrum of this source, however,
is the distinct absorption band at 1.6~$\micron$, which is
offset from 1.55--1.6~$\micron$ FeH absorption 
seen in the spectra of 2MASS~J1104+1959 (Figure~\ref{fig_nirspec})
and other mid-type L dwarfs 
\citep{cus03}.  The 1.6~$\micron$ feature is instead 
coincident with the Q-branch 
of the 2$\nu_3$ CH$_4$ band, a defining feature 
for the T dwarf spectral class.  It should be noted that
this feature appears to be
weakly present but overlooked in spectral data
from \citet{kna04}; and no mention 
is made of it by \citet{chi06}, who
also obtained SpeX prism data for {\namesh}.
Interestingly, there is no indication
of the 2.2~$\micron$ CH$_4$ band, which is commonly seen in the 
spectra of the latest-type L dwarfs (this band is 
weakly present in the spectrum
of L8/L9 2MASS~J0310+1648; Figure~\ref{fig_nirspec}).

Several of the peculiar spectral characteristics of {\namesh}
are similar to those shared
by a subclass of so-called ``blue L dwarfs'' 
\citep{cru03,cru07,kna04,me1126}, including the blue spectral energy distribution, 
strong {\water} absorption and weak CO bands.  
These properties can be explained by the presence of
thinner photospheric condensate clouds \citep{me1126}, 
which enhances the relative 
opacity of atomic and molecular species around 1~$\micron$ and 
produces bluer $J-K$ and mid-infrared colors 
\citep{mar02,kna04,leg06}.
However, \citet{gol04} have found that the thin cloud 
interpretation fails to explain the
unusually blue $K-L^{\prime}$ colors of {\namesh},
nor does it explain the presence of CH$_4$ absorption at
1.6~$\micron$ but not at 2.2~$\micron$.
Subsolar metallicity has also been cited as an explanation for the
peculiar nature of {\namesh} \citep{gol04,kna04}, although
this source does not show the extreme peculiarities
observed in the spectra of L subdwarfs \citep{me0532}, nor does subsolar
metallicity explain the presence of {\meth} absorption.

A potential clue to the nature of {\namesh} can be found by noting
that only two other late-type dwarfs have {\meth} absorption at
1.6~$\micron$ but not at 2.2~$\micron$: 2MASS~J05185995-2828372
\citep[hereafter 2MASS~J0518-2828]{cru04} and SDSS~J141530.05+572428.7
\citep{chi06}.  The latter source has not been studied
in detail, but in the case of 2MASS~J0518-2828 \citet{cru04} have found
that the combined light spectrum of an L6 plus T4 binary provides
a reasonable match to the near-infrared spectrum of this source, including
its weak {\meth} band.  Subsequent high resolution imaging has
resolved this source into two point source components 
and apparently confirm this hypothesis
\citep{mehst2}.  The similarity in the spectral peculiarities between
2MASS~J0518-2828 and {\namesh} suggests that the latter 
may be a similar but as yet unrecognized pair.

\section{Binary Template Matching}

To explore the binary hypothesis for {\namesh}, the technique of 
binary spectral template matching was employed.\footnote{For
other examples of this technique, see the analyses of \citet{mehst2,me1126,liu06,rei2252,meltbinary}; and \citet{loo07}.}
A large set of binary spectral templates was constructed from
a sample of 50 L and T dwarf SpeX prism spectra, including sources
that are unresolved in high angular resolution imaging\footnote{For an up-to-date list of known L and T dwarf binaries, see the VLM Binaries Archive maintained by Nick Siegler at \url{http://paperclip.as.arizona.edu/$\sim$nsiegler/VLM\_binaries/}.})
and are not reported as spectrally peculiar.  
The individual spectra were flux-calibrated using the 
$M_K$/spectral type relation of \citet{meltbinary}, based on published
optical and near-infrared spectral types for L dwarfs and T dwarfs,
respectively, and synthetic MKO\footnote{Mauna Kea Observatory (MKO) photometric system; \citet{sim02,tok02}.} magnitudes determined directly from the spectra.
Binaries were then constructed by combining spectral pairs
with types differing by 0.5 subclasses or more, resulting in 1164 unique 
templates.  Chi-square deviations\footnote{Here, 
$\chi^2 \equiv \sum_{\{ \lambda\} }\frac{[f_{\lambda}(0805)-f_{\lambda}(SB)]^2}{f_{\lambda}(0805)}$, where $f_{\lambda}(0805)$ is the spectrum of {\namesh}
and $f_{\lambda}(SB)$ the spectrum of the synthesized binary over the set
of wavelengths $\{ \lambda \}$ as specified in the text.}
were then computed between the spectra of
the synthesized binaries and {\namesh} over the 1.0--1.35, 1.45--1.8 and 2.0--2.35~$\micron$ regions (i.e., avoiding regions of strong
telluric absorption) after normalizing at
1.25~$\micron$. 
The single L and T dwarf spectra were also compared to that of {\namesh}
in a similar manner.

\begin{figure}
\epsscale{0.8}
\plotone{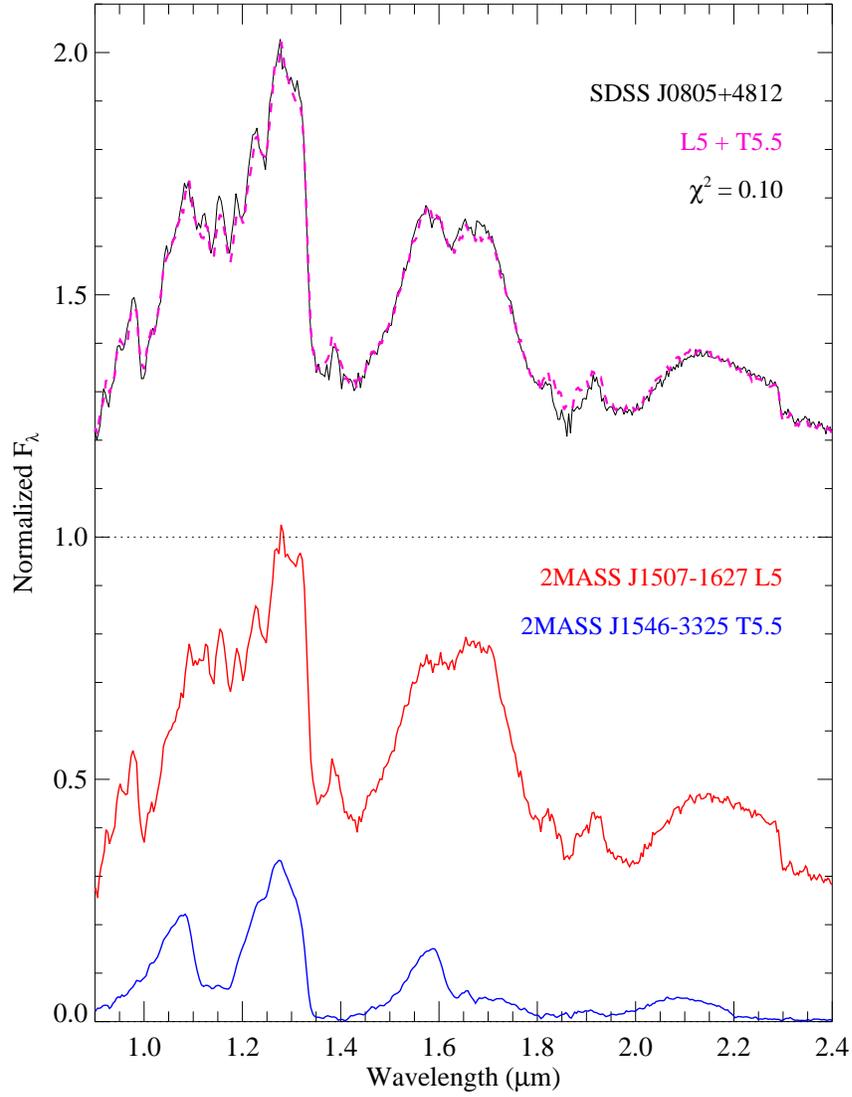}
\caption{Best match binary spectral template for {\namesh},
a combination of the L5 2MASS~J1507-1627 
and the T5.5 2MASS~J1546-3325, shown at bottom (red and blue lines, respectively).  The combined spectrum (top dashed magenta line)
is an excellent match to that of {\namesh} (top black line).  All
spectra are normalized at their 1.25~$\micron$ flux peaks, with the
spectrum of 2MASS~J1546-3325 scaled to match its relative flux 
compared to 2MASS~J1507-1627 according to the $M_K$/spectral type
relation of \citet{meltbinary}.
\label{fig_double}}
\end{figure}

The best match binary template for {\namesh}
is shown in Figure~\ref{fig_double},
composed of the L5 2MASS~J15074769-1627386 \citep[hereafter 2MASS~J1507-1627]{rei00} and the T5.5 2MASS J15462718-3325111 \citep[hereafter 2MASS~J1546-3325]{me02a}.  The combined spectrum is an excellent match to that of 
{\namesh} ($\chi^2$ = 0.10), reproducing the latter's blue spectral energy distribution,
enhanced 1.15 and 1.3~$\micron$ {\wat} absorption bands, 
weak 2.3~$\micron$ CO absorption, and
most notably the presence of weak CH$_4$ absorption at 1.6~$\micron$.
Several combinations of mid-type L dwarf and mid-type T dwarf components
produced similar excellent fits; in contrast, the
single spectral templates were all poor matches 
($\chi^2 > 1$). A mean of all binary spectral templates
with $\chi^2 < 0.5$ (33 pairs) weighted by their inverse
deviations yielded mean component types of L4.6$\pm$0.7
and T4.9$\pm$0.6.   The inferred 
primary type is notably consistent with the optical classification of {\namesh}.
This is an encouraging result, since L dwarfs are significantly brighter than
T dwarfs at optical wavelengths and should thus dominate the combined
light flux.  The inferred secondary spectral type is significantly 
later, explaining both the presence (strong absorption)
and weakness (lower relative flux) of the 
{\meth} feature at 1.6~$\micron$ in the composite spectrum of {\namesh}.
Spectral types of L4.5 and T5 are hereafter adopted for the binary components of this system.

\section{The Components of {\namesh}}

\subsection{Estimated Physical Properties}

Based on the excellent match of the spectrum of {\namesh} to empirical
binary templates composed of normal, single sources, it is compelling
to conclude that unresolved binarity provides the simplest explanation for the peculiarities of this source.  Assuming this to be the case, it is 
possible to characterize the components of {\namesh} in some detail.
Component $JHK$ magnitudes on the MKO system were determined from
reported photometry of the source \citep{kna04} and integrating
MKO filter profiles over the flux calibrated binary template spectra.
Best values, again using a weighted mean for all matches with $\chi^2 < 0.5$,
are listed in Table~\ref{tab_component}.  Comparison of the component
magnitudes to absolute magnitude/spectral type relations from 
\citet{meltbinary} yields distance estimates of 14.5$\pm$2.1~pc
and 14.8$\pm$2.5~pc
for the primary and secondary, respectively, where the uncertainties of the 
spectral types of the components and photometric magnitudes are explicitly
included.  It is of no surprise that these distance estimates 
are consistent, since the binary templates from which the component types are inferred are flux calibrated using the same absolute magnitude scales.
A mean distance of 14.6$\pm$2.2~pc is estimated for this system.

The secondary is considerably
fainter than the primary, particularly at $K$-band, where $\Delta{K}$ = 3.03$\pm$0.16 is deduced.  This suggests
a low system mass ratio ($q \equiv$ M$_2$/M$_1$).  
Using the relative $K$-band flux, $K$-band
bolometric corrections from \citet{gol04}, and assuming $q \approx 10^{-0.15\Delta{M_{bol}}}$ \citep{bur01}, $q$ = 0.48 is inferred.
This value is indeed smaller than the mass ratios of most very low-mass binaries,
77\% of which have $q \geq 0.8$ \citep{meppv}.
However, the approximation used here assumes that both components are
brown dwarfs.  
The primary is of sufficiently early type that it may 
be an older hydrogen burning low-mass star or massive
brown dwarf.  Using the evolutionary models of \citet{bur01} and 
assuming component luminosities 
calculated from the $M_{bol}$/spectral type relation of 
\citet{meltbinary},\footnote{Based on data from
\citet{gol04}.}  
estimated component masses and {\teff}s for 
ages of 1 and 5~Gyr were computed and are listed 
in Table~\ref{tab_component}.  If {\namesh}
is an older system, the mass ratio of the system increases toward unity.
This is because the slightly less massive substellar secondary has had a much
longer time to cool to T dwarf temperatures, while the primary
has settled onto the main sequence.
The strong age dependence on mass ratio estimates for low-mass stellar/substellar binaries is an important bias that is frequently overlooked.

\subsection{{\lii} Detection and Age/Mass Constraints}

\begin{figure}
\epsscale{0.8}
\plotone{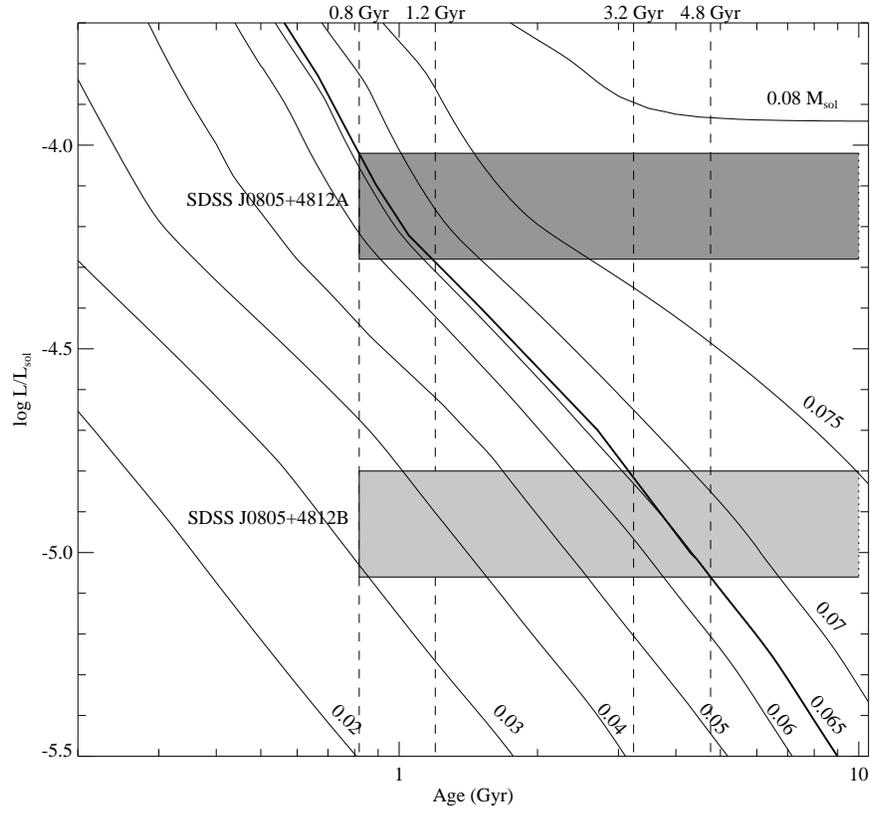}
\caption{Limits on the masses and ages of the {\namesh} components
based on their estimated luminosities (grey regions) and evolutionary
models from \citet{bur01}.  Lines trace the
evolutionary tracks for masses of 0.02 to 0.08~M$_{\sun}$.
The lithium depletion boundary is indicated by the thickened line.  
Lower age limits assuming the absence of lithium in the atmospheres of
the primary and secondary, and upper ages
limit assuming its presence, are indicated.  
The shaded regions are defined based on the absence of the 
6708~{\AA} {\lii} line in the combined light optical spectrum
of {\namesh} from \citet{haw02}.
\label{fig_evol}}
\end{figure}

From the previous discussion, it is clear that a 
robust characterization of the {\namesh} components
requires an age determination for the system, which is generally difficult
for individual field sources.  Age constraints 
may be feasible in this case, however, as
the inferred luminosities of its components straddle the {\lii} 
depletion line \citep{reb92,mag93}, as illustrated in Figure~\ref{fig_evol}.
The so-called ``binary lithium test'' pointed out by \citet{liu05} 
states that if 
lithium is present in the atmosphere of both components of the 
system, a maximum age may be inferred.  Conversely,
if lithium is absent, a minimum age may be inferred.  The
most interesting case is the absence of lithium in the primary spectrum but its
presence in the secondary spectrum, which restricts the age of the system to a finite range.

The presence of lithium in the primary may
be inferred from the 6708~{\AA} {\lii} line in 
the system's composite spectrum.  Optical data 
from \citet{haw02} show no obvious feature at this wavelength,
indicating lithium depletion in the primary and
a minimum age of 0.8~Gyr for the system based on the evolutionary models
of \citet{bur01} and the estimated component luminosities.  
However, the optical data for this faint source may 
have insufficient signal-to-noise, and the absence of the {\lii}
line requires confirmation.  For the secondary, 
the 6708~{\AA} {\lii} line is not expected to be seen
even if this component is substellar, as atomic Li is expected to be 
depleted to LiCl below temperatures of $\sim$1500~K \citep{lod99}.
In this case the presence of lithium requires
detection of the weak 15~$\micron$ band of LiCl, which has yet to be detected
in any T dwarf.  Nevertheless, if LiCl could be detected
in the spectrum of this component, the system's age could be
constrained to 1--5~Gyr.  Future observational work, perhaps
with the James Webb Space Telescope, may eventually provide the necessary
observations to make this age determination.

\subsection{{\namesh} and the L/T Transition}

\begin{figure}
\epsscale{1.1}
\plottwo{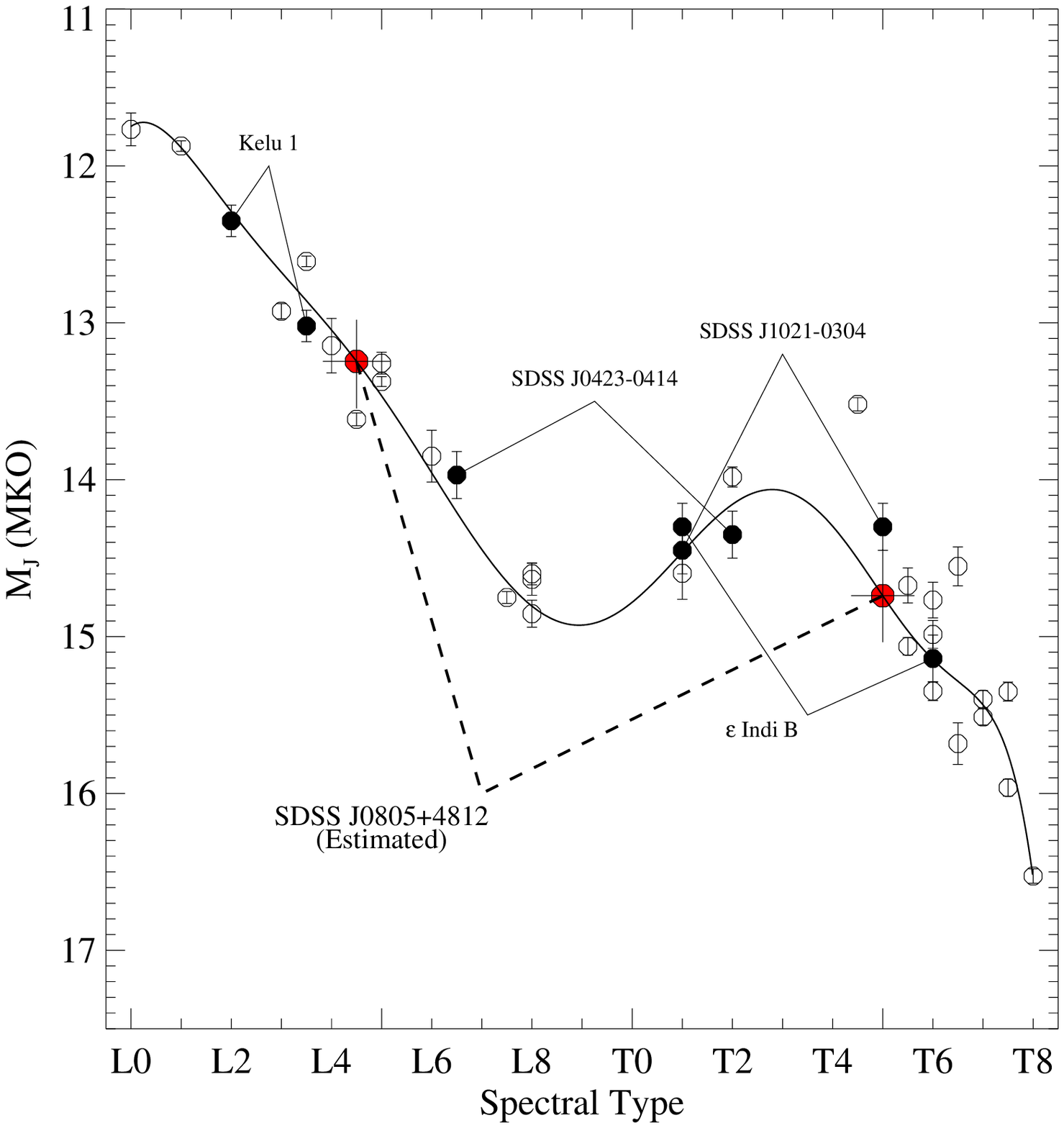}{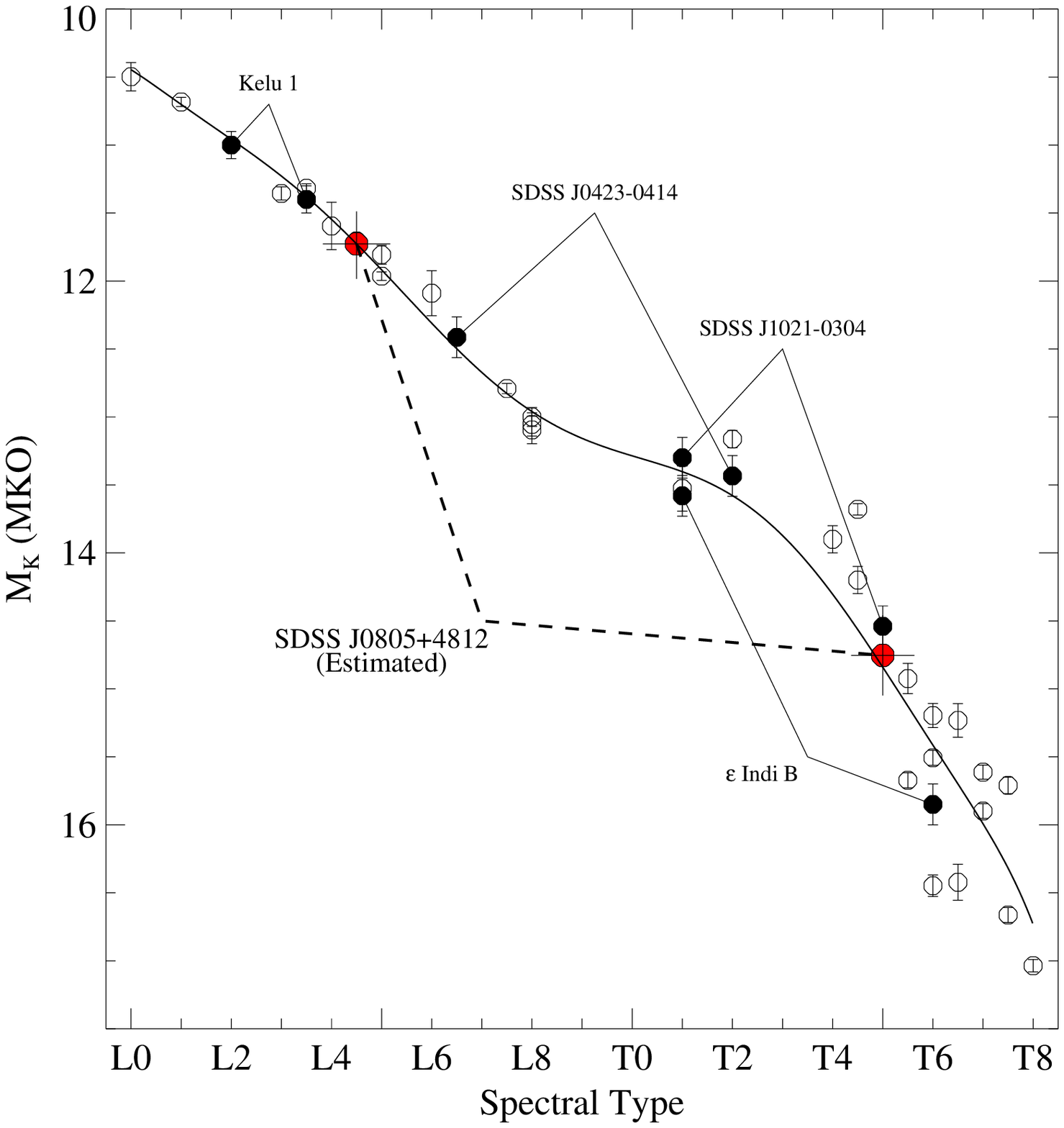}
\caption{Absolute MKO $J$ (left) and $K$ (right) magnitudes for
sources with absolute photometric errors
less than 0.2~mag.  Open circles indicate unresolved field objects, while
filled circles indicate component magnitudes for the binaries
Kelu~1AB, $\epsilon$ Indi Bab, SDSS~J0423-0414AB
and SDSS~J1021-0304AB.
Photometric data are from \citet{geb02,leg02,kna04,mcc04,liu05}; and
\citet{mehst2};
parallax data are from \citet{pry97,dah02,tin03}; and \citet{vrb04}.
The absolute magnitude/spectral type relations of \citet{meltbinary}
are delineated by thick lines.
The predicted absolute magnitudes of the {\namesh} components are
indicated by large red circles, assuming that the primary conforms
to the absolute magnitude/spectral type relations.
\label{fig_absmag}}
\end{figure}

As the inferred components of {\namesh} appear to widely straddle the L/T transition, their relative magnitudes provide a good test of
absolute magnitude/spectral type relations across this transition.   
Figure~\ref{fig_absmag} displays $M_J$ and $M_K$ magnitudes for 28 
L and T dwarfs with accurate parallax measurement ($\sigma_M \leq 0.2$~mag)
and the eight components of the binaries Kelu~1AB \citep{liu05}, 
$\epsilon$ Indi Bab \citep{mcc04}, SDSSp J042348.57-041403.5AB \citep[hereafter SDSS~J0423-0414]{geb02,me0423}
and SDSS~J102109.69-030420.1AB \citep[hereafter SDSS~J1021-0304]{leg00,mehst2}. 
To place the components of {\namesh} on this plot, the absolute magnitudes
of the primary were set to those expected from the relations from \citet{meltbinary}, which are equivalent to 
results from other studies for
mid-type L dwarfs (e.g., \citealt{tin03,kna04,vrb04,liu06}).  The absolute magnitudes of the secondary were then computed using the relative magnitudes listed in Table~\ref{tab_component}.
In both bands, there is excellent agreement between the secondary magnitudes and the absolute magnitude/spectral type relations shown. 
This is not surprising at $K$-band, since the spectral
templates used in the binary analysis were all flux-calibrated 
according to this relation.  However, the agreement at $J$-band is 
encouraging, particularly as the derived $M_J$ for {\namesh}B, 14.7$\pm$0.3, is also consistent
with absolute magnitudes for other T5-T5.5 sources.  This magnitude is also 
equivalent to values for the latest-type L dwarfs and the T1--T5 components of the resolved binaries $\epsilon$ Indi Bab, SDSS~J0423-0414
and SDSS~J1021-0304, suggesting a ``plateau'' in the
$M_J$/spectral type relation across the L/T transition, 
corresponding to a slight increase in surface fluxes at 1.05 and 1.25~$\micron$ \citep{mehst2,liu06}. However, a larger $\sim$0.5~mag brightening from types L8 to T3 
cannot yet be ruled out. It is increasingly
clear that the brightest known T dwarf, 2MASS J05591914-1404488 \citep{me0559}, with $M_J$ = 13.52$\pm$0.04 \citep{dah02,leg02}, is almost certainly a binary despite remaining unresolved
in high angular resolution observations \citep[M.\ Liu, 2007, private communication]{mehst}.

The estimated spectral types and photometric properties of the {\namesh} components are unique in that they straddle the L/T transition more widely
than other L dwarf/T dwarf binaries identified to date.  However, it is important to remember that these parameters are predictions based on the binary
spectral template analysis. Resolved photometry, radial velocity monitoring and/or parallax measurements would provide unambiguous confirmation of these results.

\section{A New Technique for Low-Mass Multiplicity Studies}

In this article, it has been demonstrated that the peculiar spectrum
of the L4/L9.5 {\namesh} can be sufficiently explained as the combined
light spectrum of an L4.5 plus T5 unresolved binary.  
This source joins a growing list of L/T transition binaries, many of
which exhibit the same spectral peculiarities as {\namesh}
(blue near-infrared colors, presence
of 1.6~$\micron$ {\meth} absorption without the 2.2~$\micron$ band) and 
have been subsequently resolved through high resolution imaging
\citep{cru04,me0423,mehst2,liu06,rei2252}.  
The similarity in the spectral 
peculiarities of these sources 
suggests that other binaries
composed of L and T dwarf components could be readily identified 
and characterized through analysis of low-resolution,
combined-light, near-infrared spectroscopy alone, as has been demonstrated
here.  This is a promising 
prospect, as traditional high resolution imaging or spectroscopic
techniques are limited by resolution and geometry restrictions,
such that closely-separated binaries and/or distant binary
systems can be overlooked. This is particularly
a concern for brown dwarf binaries,
over 80\% of which 
have projected separations less than 20~AU \citep{meppv}.
The use of combined
light spectra in binary studies are not subject to resolution limitations,
enabling the identification of binaries independent of
separation.  Furthermore, low resolution near-infrared spectroscopy
is far less resource intensive than high resolution imaging and spectroscopic
techniques, which requirie the use of large telescopes
and/or space-based platforms.

On the other hand, spectral peculiarities arising from binaries
will arise only over a limited range of mass ratios (e.g., they will not generally
be seen in equal-mass systems), may be readily apparent only in systems
composed of L dwarf plus T dwarf components, and must be distinguished from 
spectral peculiarities arising from other effects such as metallicity, surface gravity
or condensate cloud structure.  Hence, the phase space in which unresolved
binaries may be identified via low-resolution spectroscopy may be 
restricted in a non-trivial way, and its characterization is beyond
the scope of this study.
Nevertheless, the results presented here 
should make it clear that low-resolution
near-infrared spectroscopic analysis
provides a complementary approach to traditional high resolution
imaging and spectroscopic techniques in the identification
and characterization of low-mass stellar and substellar binaries.

\acknowledgements

The author would like to thank telescope operator Dave Griep
and instrument specialist John Rayner at IRTF
for their assistance during the
observations, and the anonymous referee for her/his helpful
critique of the original manuscript.  This publication makes
use of data from the Two Micron All Sky Survey, which is a joint
project of the University of Massachusetts and the Infrared
Processing and Analysis Center, and funded by the National
Aeronautics and Space Administration and the National Science
Foundation. 2MASS data were obtained from the NASA/IPAC Infrared
Science Archive, which is operated by the Jet Propulsion
Laboratory, California Institute of Technology, under contract
with the National Aeronautics and Space Administration.
This research has benefitted from the M, L, and T dwarf compendium housed at DwarfArchives.org and maintained by Chris Gelino, Davy Kirkpatrick, and Adam Burgasser.
The authors wish to recognize and acknowledge the 
very significant cultural role and reverence that 
the summit of Mauna Kea has always had within the 
indigenous Hawaiian community.  We are most fortunate 
to have the opportunity to conduct observations from this mountain.

Facilities: \facility{IRTF~(SpeX)}

\clearpage

\clearpage

\clearpage

\clearpage

\clearpage

\begin{deluxetable}{lccc}
\tabletypesize{\footnotesize}
\tablecaption{Predicted Component Parameters for {\namesh}AB. \label{tab_component}}
\tablewidth{0pt}
\tablehead{
\colhead{Parameter} &
\colhead{{\namesh}A} &
\colhead{{\namesh}B} &
\colhead{Difference}  \\
}
\startdata
Spectral Type & L4.5$\pm$0.7 & T5$\pm$0.6 & \nodata \\
${J}$\tablenotemark{a} & 14.25$\pm$0.04  &  15.75$\pm$0.08 & 1.50$\pm$0.09 \\
${H}$\tablenotemark{a} & 13.62$\pm$0.03  &  16.01$\pm$0.14 & 2.39$\pm$0.15 \\
${K}$\tablenotemark{a} & 12.37$\pm$0.03  &  15.40$\pm$0.16 & 3.03$\pm$0.16 \\
$\log_{10}{L_{bol}/L_{\sun}}$\tablenotemark{b} & -4.15$\pm$0.13 & -4.93$\pm$0.13 & 0.88$\pm$0.16 \\
Mass (M$_{\sun}$) at 1 Gyr\tablenotemark{d} & 0.066 & 0.036 &  0.55\tablenotemark{c} \\  
Mass (M$_{\sun}$) at 5 Gyr\tablenotemark{d} & 0.078 & 0.069 &  0.88\tablenotemark{c} \\
{\teff} (K) at 1 Gyr\tablenotemark{d} & 1830$\pm$90 & 1200$\pm$70 &  \nodata \\ 
{\teff} (K) at 5 Gyr\tablenotemark{d} & 1780$\pm$100 & 1100$\pm$70 &  \nodata \\ 
Estimated $d$ (pc) & 14.5$\pm$2.1 & 14.8$\pm$2.5 & -0.3$\pm$0.5 \\
\enddata
\tablenotetext{a}{Synthetic magnitudes on the MKO system.}
\tablenotetext{b}{{L}uminosities based on the $M_{bol}$/spectral type relation of \citet{meltbinary}.}
\tablenotetext{c}{Mass ratio (M$_2$/M$_1$).}
\tablenotetext{d}{Based on the evolutionary models of \citet{bur01} and the estimated luminosities.}

\end{deluxetable}

\end{document}